# STATISTICAL METHODS FOR AUTOMATED DRUG SUSCEPTIBILITY TESTING: BAYESIAN MINIMUM INHIBITORY CONCENTRATION PREDICTION FROM GROWTH CURVES

By Xi Kathy Zhou,[1,2] Merlise A. Clyde,[1,3] James Garrett, Viridiana Lourdes,[1] Michael O'Connell, Giovanni Parmigiani,[1] David J. Turner and Tim Wiles

*Cornell University, Duke University, Becton–Dickinson Diagnostic Systems, Morgan Stanley, Insightful Corporation, Johns Hopkins University, Becton–Dickinson Diagnostic Systems and Becton–Dickinson Diagnostic Systems*

Determination of the minimum inhibitory concentration (MIC) of a drug that prevents microbial growth is an important step for managing patients with infections. In this paper we present a novel probabilistic approach that accurately estimates MICs based on a panel of multiple curves reflecting features of bacterial growth. We develop a probabilistic model for determining whether a given dilution of an antimicrobial agent is the MIC given features of the growth curves over time. Because of the potentially large collection of features, we utilize Bayesian model selection to narrow the collection of predictors to the most important variables. In addition to point estimates of MICs, we are able to provide posterior probabilities that each dilution is the MIC based on the observed growth curves. The methods are easily automated and have been incorporated into the Becton–Dickinson PHOENIX automated susceptibility system that rapidly and accurately classifies the resistance of a large number of microorganisms in clinical samples. Over seventy-five studies to date have shown this new method provides improved estimation of MICs over existing approaches.

Received January 2008; revised October 2008.

[1]Supported in part by Becton–Dickinson.

[2]Supported in part by the Clinical and Translation Science Center at Weill Cornell Medical College through the National Institute of Health Grant UL1 RR024996.

[3]Supported in part by NFS Grant DMS-04-06115. Any opinions, findings and conclusions or recommendations expressed in this material are those of the authors and do not necessarily reflect the views of the National Science Foundation.

**1. Introduction.** Since the discovery of penicillin in the late 19th century, microbiology has undergone rapid development. Large numbers of antibiotics have been identified and have greatly improved the management of patients with infectious diseases. Antimicrobial susceptibility testing (AST) of clinically obtained isolates[4] is performed daily across the world. Through such tests, the activity of an antimicrobial agent against an organism, such as bacteria, is reported either quantitatively or qualitatively. Quantitative microbial inhibitory activity is typically described in terms of the minimum inhibitory concentration (MIC), which is defined as the lowest concentration of an antimicrobial agent at which bacterial growth is inhibited in in-vitro testing. In the U.S. the Clinical and Laboratory Standards Institute [CLSI (2008)] establish MIC breakpoints for a given bacterial species and antibiotic so that an isolate may be classified as "susceptible" (S) if the MIC is less than or equal to the lower breakpoint, "resistant" (R) if the MIC is greater than or equal to the higher breakpoint, or "intermediate" (I) if the MIC falls in between. Isolates in the susceptible category are inhibited by the usually achievable concentrations of antimicrobial agent when the recommended dosage is used for the site of infection, while isolates in the resistant group are not inhibited by the usually achievable concentrations of the agent with normal dosage schedules; the intermediate classification provides a buffer zone for uncontrolled variation or where the antibiotic may be effective at higher than normal doses. Results of such testing are then used to predict treatment outcome with the antimicrobial agents tested and guide clinicians in selecting the most appropriate agent for a particular clinical problem [Turnidge, Ferraro and Jorgensen (2003)].

Current reference methods for antimicrobial susceptibility testing, such as dilution or disk diffusion, are predominantly phenotypic approaches based on growth patterns of micro-organisms in antimicrobial agents [Wheat (2001)]. Dilution susceptibility testing methods are quantitative methods that determine the MIC by exposing the isolate to a series of two-fold dilutions of the antimicrobial agent (e.g., 1 $\mu$g/mL, 2 $\mu$g/mL, 4 $\mu$g/mL, 8 $\mu$g/mL, 16 $\mu$g/mL, etc.) in a suitable culture system, such as broth or agar, on a series of plates or tubes. Typically, eight or more concentrations of an agent are used, although the range of concentrations tested depends on the antimicrobial agent and species being tested. These are incubated overnight and each tube or plate is visually inspected for growth; the lowest dilution in which the isolate does not grow is reported as the MIC [see Jorgensen and Turnidge (2003), CLSI (2006) for detailed discussion and methods]. Including preparation time, incubation time and analysis, it may take 24–48 hours to provide results.

---

[4]A population of bacteria or other cells that has been isolated.

In recent years, efforts for speeding up the process of susceptibility testing have resulted in the development of automated AST systems [Ferraro and Jorgensen (2003)]. In an automated system, it is feasible to monitor the growth of isolates in real time; by utilizing the information in the growth curves corresponding to the series of dilutions, it is possible to make an early determination of the MIC (in a couple of hours ideally), providing a dramatic decrease in the time to results compared to reference methods that make a determination at a fixed end-point. Given the critical importance of rapid and accurate identification of micro-organisms and especially their drug resistance in the management of patients, automation of the testing process and rapid reporting of results is of great clinical and financial benefit to patients and hospitals [Barenfanger, Drake and Kacich (1999)]. With automation, a large number of tests can be performed simultaneously, leading to a substantial increase in the amount of data to be analyzed. Thus, there is a crucial need for such automated systems to be coupled with rigorous statistical methods that produce reliable MIC estimates in real-time. In this paper we develop a novel statistical method that allows accurate and rapid determination of MICs based on a panel of growth curves for a given isolate exposed to various dilutions of an antimicrobial agent. This methodology is currently implemented in Becton–Dickinson's (BD) newly developed PHOENIX AST system (BD Diagnostic Systems, Sparks, MD). This system provides rapid and reliable susceptibility testing of a majority of clinically encountered bacterial strains and is currently utilized by a large number of laboratories across the globe.

In the next section we describe the structure of the panel growth data obtained during the development of the BD PHOENIX AST system. Our goal is to identify and use features of the growth curves to accurately predict the MIC for each set of panel data. In order to train the statistical algorithms, in Section 3 we develop a model that predicts for each concentration a growth or no growth response, using available features of the curves from each panel. Features in the growth data that are crucial for predicting the probability of growth are selected using Bayesian model selection. For each isolate and drug dilution in a panel, we construct an estimate of the probability of growth using these selected features of the growth curves. In Section 4 we present a novel method to combine the estimated growth probabilities for the sequence of drug dilutions in a panel to construct a probability distribution for the MIC. A decision theoretic approach for estimating the MIC is presented which balances the different types of errors in making an MIC call. In Section 5 we validate the method and illustrate its performance using two selected antibiotics, piperacillin (PIP) and cefoxitin (FOX). We conclude with discussion in Section 6.

**2. The PHOENIX AST system.** PHOENIX is an instrumented antimicrobial susceptibility testing (AST) system for rapid identification and susceptibility determination of bacterial isolates from clinical samples. In this system a bacterial suspension is placed in a series of wells (typically eight) on a test panel, where each panel is a single PHOENIX diagnostic disposable cartridge. Each well in a panel corresponds to a different dilution of the microbial agent being tested. Multiple panels are placed in a revolving carousel and at a sequence of twenty minute intervals the system moves the panels past a detector where red, green and blue wavelengths are directed at each well. The system measures the wavelengths that emanate from the wells; these optical measurements are used to generate two measures of microbial growth: the redox state and turbidity characteristics of the sample wells which are directly correlated with microbial growth. The redox state of a sample in a well is measured by utilizing the change in red, green and blue readings that occurs over time as a result of the reduction of a growth indicator, such as resazurin, by the microbial material in the well. As the resazurin is reduced, the color of the sample in the well changes from blue to red to clear. This change in redox is represented numerically as a continuum, with the value 0 indicating an unreduced growth indicator (blue = resazurin), the value 0.5 indicating that the indicator is 50% reduced (red), and the value 1.0 indicating that the indicator has been completely reduced (clear = dihydroresorufin). The turbidity is also estimated by using the red, green and blue readings. The initial signal has a value of 0 and a maximum of 2.25 (McFarland) units can be estimated.

The system monitors growth over time to decide if the samples have incubated long enough to estimate the MIC. If the processor determines that the maximum redox state for the growth control well of the test panel is greater than 0.07 but less than a predetermined value of 0.2, the panel continues to be incubated, as this implies insufficient growth. If the processor determines that the maximum redox state for the growth control well is indeed greater than 0.2, the processor then checks whether a call can be made. If the processor determines that the redox curve for the growth control well indicates that the sample is not yet classifiable as either a slow or fast growing sample, the panel incubation continues and the panel is retested twenty minutes later. If the processor determines that the redox curve classification is either fast or slow growing, the time at which this occurs is labeled as the "time-to-result" and the turbidity and redox data for each of the wells in the panel are extracted. If the incubation period exceeds 16 hours, the test panel is reported as "failed" for having insufficient sample growth in the allotted time, and no results are reported for that test panel. A second-degree polynomial local regression algorithm (LOESS) is used to smooth the time series data for both redox and turbidity values calculated for each well over the elapsed period of time. From the LOESS fit any reading at any time point can be estimated. Moreover, first- and second-derivatives can be estimated at any point as a function of fitted local regression coefficients.

Figure 1 illustrates these interpolated growth curves for three PHOENIX panels in which three gram-negative bacterial strains were combined with the following dilutions of the drug piperacillin (PIP): 0.25 $\mu$g/mL, 0.5 $\mu$g/mL, 1 $\mu$g/mL, 2 $\mu$g/mL, 4 $\mu$g/mL, 8 $\mu$g/mL, 16 $\mu$g/mL, 32 $\mu$g/mL, 64 $\mu$g/mL and 128 $\mu$g/mL.

Each row of plots shows the growth response of a single isolate to varying dilutions of PIP, and was selected to show the range of responses. For a given plot, each curve corresponds to a different dilution of the drug PIP. The vertical line represents the time-to-result, the time at which features are extracted for determining the MIC. Independently from the PHOENIX AST system, a reference MIC is determined for each strain using broth micro-dilution susceptibility testing with the same series of two-fold dilutions [CLSI (2006)]. In Figure 1 curves corresponding to dilutions less than the reference MIC are dashed lines, while all curves with dilutions greater than the reference MIC are solid. The dilution corresponding to the reference MIC is represented with a long and short dashed line.

**3. Modeling probability of growth.** Theoretically, we should see no growth in a well with a concentration higher than the true MIC. Using the external reference value of the MIC, we create a binary response variable $Y_{ij}$ for dilution $j$ in panel $i$ defined as

$$Y_{ij} = \begin{cases} 1, & \text{if dilution } j \text{ is } < \text{ reference MIC,} \\ 0, & \text{if dilution } j \text{ is } \geq \text{ reference MIC.} \end{cases}$$

This definition treats the reference MIC data as if it were measured without error and ignores any possible missclassifications of curves. The reference methods are discrete measurements of what is in reality a value along a continuum. The error associated with any individual MIC estimate will be affected by many factors; these include but are not limited to operator error, variation in the materials utilized and the isolate and antimicrobial agent being tested. Multiple observations with a specific bacterial isolate and antimicrobial combination will typically be normally distributed around a modal MIC value, with 60–80% of the observations being at the modal MIC value [CLSI (2006)]. Conventionally, any two observations within $\pm$ one two-fold dilution are considered to be in agreement (this corresponds to a one unit change on the log base 2 scale). For our purposes, we will treat the reference MIC estimate as the truth, as this is the best available information.

Inspection of the panels in the third row of Figure 1 indicates that several curves with dilutions above the reference MIC exhibit growth in both the redox and turbidity responses for strain 14,617. For strain 14,598 (the second row), the redox curves also show evidence of growth at dilutions above the reference MIC; the turbidity measurements for dilutions above the MIC, while still increasing, are well separated from the curves above the MIC. Upon further verification, the problem for these two isolates is not attributed to an error in the reference MIC data, but attributed to the PHOENIX measurements (also subject to operator error, variation in materials such as formulation of the panels[5] and variation in the isolate). The predictive accuracy of the training model may be improved through an iterative procedure by omitting such "outlier" panels that exhibited strong growth for dilutions above the MIC when the reference MIC could be verified as being correct. Because such iterative "outlier" analysis could not be conducted for model training with additional agents tested at BD, we chose to include such cases in our analyses.

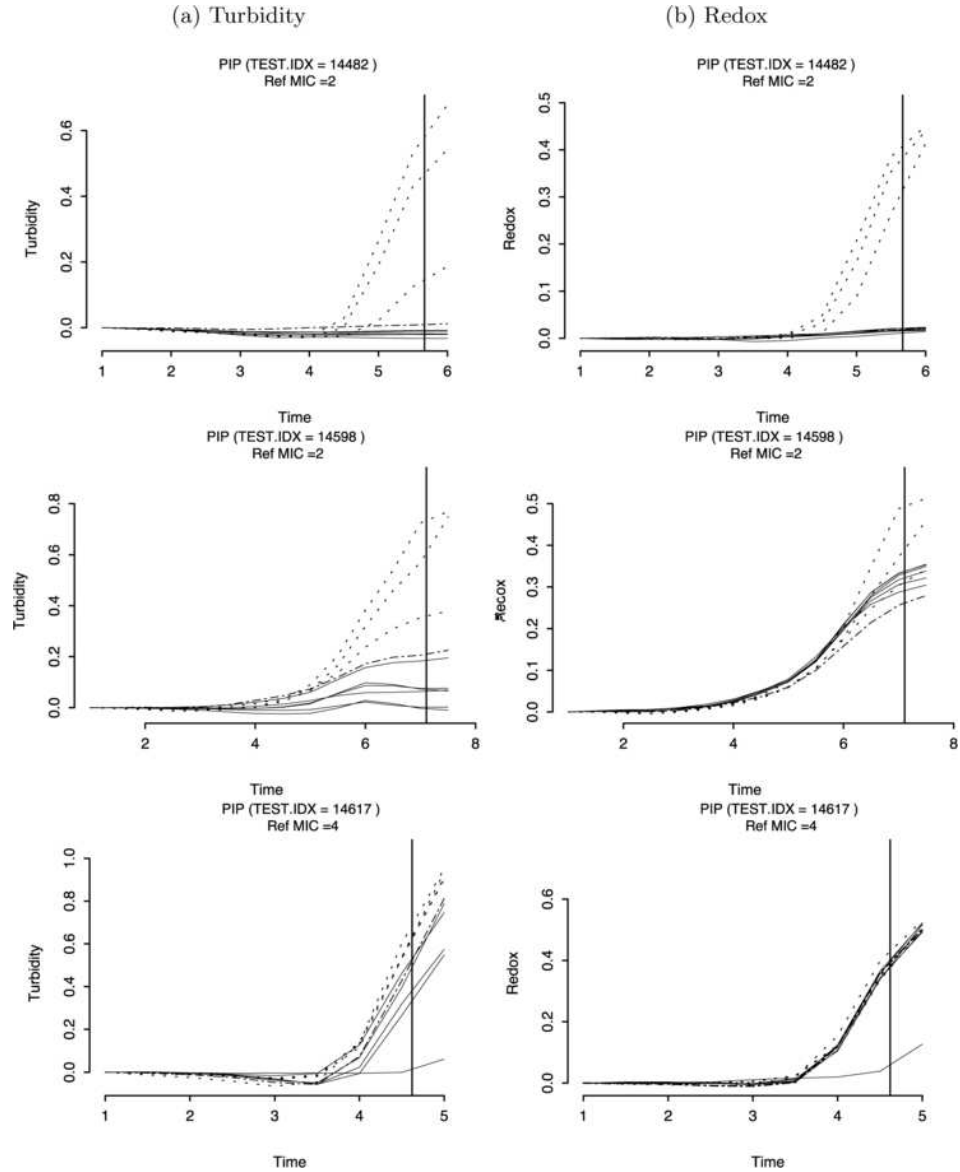

FIG. 1. *Panel data for three gram-negative bacterial strains exposed to dilutions of piperacillin (PIP) showing turbidity (a) and redox (b) over time in hours. The vertical line is at the time-to-result when features are extracted for estimating the MIC. Dashed curves have concentrations less than the reference MIC, while solid lines have concentrations greater than or equal to the reference MIC. The curve with dilution equal to the reference MIC is depicted with long–short dashes.*

3.1. *Models.* Conditional on the unknown probabilities (and MIC), we assume that indicators of growth in different wells and across panels are independent of each other,

$$(3.1) \qquad Y_{ij} \mid \pi_{ij} \stackrel{\text{ind}}{\backsim} \text{Ber}(\pi_{ij}), \qquad i=1,\ldots,n, \ j=1,\ldots,J,$$

where $\pi_{ij}$ is the probability of growth or, equivalently, that dilution $j$ in panel $i$ is less than the MIC, $n$ is the number of panels, and $J$ is the number of dilutions or wells in a panel. While it is conceivable that observations in the same panel may be dependent, the model in (3.1) implies that such dependence may be captured through the model for the probabilities $\pi_{ij}$, for example, a panel specific random effect.

The indicator of GROWTH/NO GROWTH may be predicted as a function of the growth curves by modeling the probabilities of growth, $\pi_{ij}$, as a function of the time series for dilution $j$ in panel $i$. While of course the entire time series could be used for modeling the GROWTH indicators, certain features of the curves may be sufficient for differentiating the pattern of GROWTH/NO GROWTH. In an experiment with no noise in the growth curves, we would actually need only the difference in growth measurements from the beginning to the time-to-result to separate the dilutions corresponding to GROWTH or NO GROWTH. However, with noisy curves, other features may be better predictors of growth. In Table 1 we list the features of the curves that were viewed as being scientifically relevant for predicting growth. These include the difference in growth, the area under the growth curve, the first derivative, the second derivative at the time-to-result and time points where maxima occurred. Most features are defined relative to the growth control well to standardize them across different isolate/drug combinations.

---

[5]The experimental data presented here were from the early stages of development of the PHOENIX system; BD has reformulated panels since our initial analysis was conducted.

TABLE 1
*Characteristic features of growth curves used as covariates in the GROWTH or NO GROWTH model*

| Feature label | Description |
| --- | --- |
| T.FD | Turbidity, 1st derivative of the Test Well |
| T.SD | Turbidity, 2nd derivative of the Test Well |
| T.IN | Turbidity, Integral of the Test Well |
| T.AB.M | Turbidity, Maximum Absolute of the Test Well |
| T.FD.M | Turbidity, Maximum 1st Derivative of the Test Well |
| T.SD.M | Turbidity, Maximum 2nd Derivative of the Test Well |
| T.AB.M.R | T.AB.M(Test Well)/T.AB.M(Control Well) |
| T.FD.M.R | T.FD.M(Test Well)/T.FD.M(Control Well) |
| T.SD.M.R | T.SD.M(Test Well)/T.SD.M(Control Well) |
| T.IN.R | T.IN(Test Well)/T.AB(Control Well) |
| T.FD.T | $t_{\text{T.FD.M}}$(Test Well)$-t_{\text{T.FD.M}}$(Control Well) |
| T.SD.T | $t_{\text{T.SD.M}}$(Test Well)$-t_{\text{T.SD.M}}$(Control Well) |
| R.FD | Redox, 1st derivative of the Test Well |
| R.SD | Redox, 2nd derivative of the Test Well |
| R.IN | Redox, Integral of the Test Well |
| R.AB.M | Redox, Maximum Absolute of the Test Well |
| R.FD.M | Redox, Maximum 1st Derivative of the Test Well |
| R.SD.M | Redox, Maximum 2nd Derivative of the Test Well |
| R.AB.M.R | R.AB.M(Test Well)/R.AB.M(Control Well) |
| R.FD.M.R | R.FD.M(Test Well)/R.FD.M(Control Well) |
| R.SD.M.R | R.SD.M(Test Well)/R.SD.M(Control Well) |
| R.IN.R | R.IN(Test Well)/R.AB(Control Well) |
| R.FD.T | $t_{\text{R.FD.M}}$(Test Well)$-t_{\text{R.FD.M}}$(Control Well) |
| R.SD.T | $t_{\text{R.SD.M}}$(Test Well)$-t_{\text{R.SD.M}}$(Control Well) |

We use a logistic regression model, a generalized linear model, to relate the GROWTH indicator to the collection of features, where the linear predictor $\eta_{ij} \equiv \text{logit}(\pi_{ij}) \equiv \log(\pi_{ij}/(1-\pi_{ij}))$ is expressed as a linear function of selected features. As we are uncertain that the relationship is actually linear in the features in Table 1, we may try more flexible generalized linear models where we replace the linear functions by up to a third order polynomial in each feature. As the linear, quadratic and cubic terms are typically highly correlated with each other, orthogonal polynomials are preferable from a computational perspective for model selection, but give equivalent predictions. More generally, the generalized linear model can be represented in matrix notation as $\boldsymbol{\eta} = \mathbf{X}\boldsymbol{\beta}$, where $\mathbf{X}$ represents the $n \times p$ matrix of feature variables with columns for the linear, quadratic and cubic terms, $\boldsymbol{\beta}$ is the $p$ dimensional vector of the unknown regression coefficients and $\boldsymbol{\eta}$ is the $n$-dimensional vector of the linear predictor. Once we have estimates of $\boldsymbol{\beta}$, these are used to calculate $\hat{\eta}_{ij}$, the estimate of the linear predictor, which in turn is used to obtain the estimates of the probabilities of GROWTH,

$$\hat{\pi}_{ij} = \frac{\exp(\hat{\eta}_{ij})}{1 + \exp(\hat{\eta}_{ij})}.$$

We may find that not all features are needed to model the probability of GROWTH; the variables represent features that could potentially explain the growth patterns and in some cases may be redundant. Using all features (and the higher polynomial terms) may lead to over-fitting of the model training data and poor predictions on the out-of-sample validation data. Reducing the set of features is also important because of limited storage and computing capacity in the system for making predictions. Bayesian variable selection [Hoeting et al. (1999), Clyde and George (2004)] can be used to reduce the set of features to prevent over-fitting while still providing excellent out-of-sample properties.

3.2. *Bayesian model selection.* We used Bayesian model selection based on the Bayes Information Criterion [Schwarz (1978), Kass and Raftery (1995)] to reduce the set of features used in the prediction of GROWTH. This approach is equivalent to using a penalized deviance criterion for model selection, however, the penalty depends on the sample size, which ensures that in large samples that the probability of the true model goes to one (under modest conditions), and provides consistent model selection [Kass and Raftery (1995)]. This choice typically results in more parsimonious models than selection based on, for example, the often-used Akaike's Information Criterion (AIC) [Akaike (1973, 1983a, 1983b)]. Hoeting et al. (1999) provide examples where the use of BIC in Bayesian model selection and model averaging leads to excellent predictive performance in practice.

Models $\mathcal{M}$ correspond to different choices of features and polynomials in the features used to capture the smooth functions. Using BIC, the probability of a model, $\mathcal{M}_m$, given the collection of panel data $\mathbf{Y} = \{Y_{ij}\}_{i=1,\ldots,n,j=1,\ldots,J}$ is approximated by

$$p(\mathcal{M}_m|\mathbf{Y}) = \frac{\exp(\mathcal{L}(\mathcal{M}_m) - (d_m/2)\log(n))}{\sum_{m'} \exp(\mathcal{L}(\mathcal{M}_{m'}) - (d_{m'}/2)\log(n))},$$

where $\mathcal{L}(\mathcal{M}_m)$ is the log likelihood under model $\mathcal{M}_m$ evaluated at the maximum likelihood estimates $\hat{\beta}$ and $d_m$ is the number of terms in the model. Because of the large number of features and higher order terms, we cannot enumerate all models. To identify the high posterior probability models, we used the deterministic search algorithm `bic.glm()` in S-PLUS [Hoeting et al. (1999)].

Because of the large number of potential features and limitations in the leaps and bounds algorithm used in `bic.glm()`, it was necessary to use a two stage procedure to identify which features would be incorporated in the selected model. The first stage was a screening stage where we identified the highest probability models that included only linear terms in the features. Using the subset of features that were included in the top model, we then added quadratic and cubic terms in these features, and repeated the calculations of the posterior model probabilities. While such a sequential approach may miss the highest probability model,[6] this scheme corresponds to a hierarchical model that incorporates second order (and higher) terms only if there are important main (linear) terms. Second, because the search algorithm needed to be run for hundreds of other drugs, this led to a reasonably efficient computational strategy. After this two stage selection approach, the highest posterior probability model was used to determine the distribution of the MIC, as described in Section 4.

The above procedures are applied to study the turbidity and redox growth curves for gram-negative bacterial strains exposed to dilutions of cefoxitin (FOX) ($n = 1647$ panels) and piperacillin (PIP) ($n = 1599$ panels). These two antibiotics were selected to provide a testbed for method development. FOX was considered to be an easier system to model, with more clearly distinguished cases of GROWTH/NO GROWTH, while PIP was viewed as being a more challenging case, as illustrated with the selected growth curves in Figure 1. For modeling purposes here, both fast and slow growing strains were combined. The data for FOX were obtained using the following series of two-fold dilutions $\{2^{-1}, 2^0, 2^1, \ldots, 2^5\}$ ($J = 7$), while the data for PIP used $\{2^{-2}, 2^{-1}, \ldots, 2^7\}$ ($J = 10$).

The selected features are dependent on the drug. The model for FOX includes five terms: `T.FD.M.R` the ratio of the turbidity maximum first derivative of the test well to the turbidity maximum first derivative of the control well; `R.FD` first derivative of redox in the test well; `R.SD` second derivative of redox in the test well; `R.SD.M` maximum second derivative of redox in the test well; and `R.AB.M.R` the ratio of the redox maximum absolute value of the test well to the redox maximum absolute value of the control well. The model for the more challenging PIP data includes nine terms: an `T.AB.M.R` the ratio of the turbidity maximum absolute value of the test well to the turbidity maximum absolute value of the control well; `T.FD` first derivative of turbidity in the test well; `T.SD` second derivative of turbidity in the test well; `T.FD.M.R` ratio of absolute first derivative of turbidity of the test well to maximum first derivative of turbidity in the control well; `T.SD.M` maximum second derivative of turbidity in the test well; `R.AB.M.R` the ratio of the redox maximum absolute value of the test well to the redox maximum absolute value of the control well; `R.FD` first derivative of redox in the test well; R.FD.M.R ratio of absolute first derivative of redox of the test well to maximum first derivative of redox in the control well; and R.SD second derivative of redox in the test well.

---

[6] If there are strong quadratic or cubic effects, but the linear terms are not important, then we may fail to select this model with the sequential approach.

In modeling GROWTH/NO GROWTH, we made no provision to order the estimated probabilities $\pi_{ij}$ in panel $i$ according to dilution. Rather than explicitly building an order constraint into the model for GROWTH/NO GROWTH, we take the theoretical ordering into account in developing the prediction of the MIC, as described in the next section.

## 4. MIC estimation.

4.1. *Overview of MIC estimation.* The modeling described in Section 3 operates at the level of a single dilution or well in a panel, and is aimed at determining the probability that each dilution is displaying growth. Note, however, that prediction of the MIC must be done at the level of a panel. In this section we describe our strategy for combining the dilution-level predictions and obtaining a MIC estimate for each panel. For panel $i$, our prediction algorithm consists of the following steps:

*Estimate growth probabilities.* For each dilution $j$ in panel $i$ estimate the probability $\pi_{ij}$ that there is growth, $\hat{\pi}_{ij}$.

*Combine growth probabilities.* For each dilution $j$ in panel $i$ estimate the probability $\rho_{ij}$ that dilution $j$ is the MIC. This distribution of the MIC is constructed by combining all the $\hat{\pi}_{ij}$ for panel $i$ and is described in detail in Section 4.2.

*Estimate the MIC.* Using the distribution of the MIC in panel $i$, $\rho_{ij}$, derive an estimate for the MIC for the panel. Two estimates are discussed in detail: the "modal MIC" and a "decision theoretic MIC." The probability distribution of the MIC may also be used to delay the call, when there is a high degree of uncertainty about which dilution is the MIC.

Statistically, this overall procedure achieves the important practical goals of (a) estimating growth based on simple physical features representing growth in a well; (b) training the growth probability model on large data sets; (c) adjusting the probability of growth in a well depending on the observations made on other wells in the same panel; and (d) providing the basis for sequential estimation of the MIC, and decisions about delaying the call.

4.2. *Probability distribution of the MIC.* We now describe how to derive the probabilities $\rho_{i1}, \ldots, \rho_{iJ+1}$ that dilution $j$ is the MIC of the $i$th panel, using the set of probabilities $\hat{\pi}_{i1}, \ldots, \hat{\pi}_{iJ}$ that there is growth in each well. Suppose that the dilution for well $j$ (denoted as $D_j$) is the MIC for panel $i$. If this were true, theoretically we would have the following sequence of curves: all curves with dilutions less than $D_j$ would exhibit GROWTH, and all dilutions greater than or equal to $D_j$ would have NO GROWTH. How do we compute the probability of this sequence given the probability of GROWTH/NO GROWTH for each well?

First, we make additional assumptions of order (O) and independence (I):

O: Suppose that the dilution (concentration) in well $j$, $D_j$, is greater than the dilution in well $k$, $D_k$. Then, if $D_k$ is inhibitory, so is $D_j$.

I: Conditional on the set of probabilities $\pi_{ij}$ for $j = 1, \ldots, J$, the outcome in well $j$, $Y_{ij}$ is independent of the outcome $Y_{ik}$ in any other well $k \neq j$.

We utilized Assumption I in constructing the model for GROWTH/NO GROWTH. Within the nested restrictions of Assumption O, the probability of the sequence of curves that can lead to $D_j$ being the MIC can be computed under independence, Assumption I.

Consider the set of all sequences of growth curves that satisfy the ordering Assumption O. We call these "valid" sequences. For example, if a panel had only three wells, the set of valid sequences of GROWTH/NO GROWTH results would be

| $D_1$ | < | $D_2$ | < | $D_3$ | Call |
|---|---|---|---|---|---|
| 0 | | 0 | | 0 | $MIC \leq D_1$ |
| 1 | | 0 | | 0 | $D_1 < MIC \leq D_2$ |
| 1 | | 1 | | 0 | $D_2 < MIC \leq D_3$ |
| 1 | | 1 | | 1 | $MIC > D_3$ |

Note that while there are $J$ dilutions, there are $J+1$ valid sequences, since if all $J$ observed curves exhibit growth, the call should be that the MIC is greater than the highest observed dilution.

By restricting attention to the reduced set of ordered sequences, we are now in a position to compute the probability $\rho_{ij}$ of the sequence of curves that leads to $D_j$ being the MIC in panel $i$. This ordered sequence is

| $D_1$ | $\ldots$ | $D_{j-1}$ | $D_j$ | $\ldots$ | $D_J$ |
|---|---|---|---|---|---|
| 1 | 1 | 1 | 0 | 0 | 0 |

By comparing the likelihood of the sequence above to the likelihood of the other valid sequences, we can compute MIC probabilities that are consistent with the ordering Assumption O,

$$(4.1) \quad \rho_{ij} = P(D_{j-1} < MIC_i \leq D_j) = \frac{\prod_{k<j} \pi_{ik} \prod_{k \geq j}(1 - \pi_{ik})}{\sum_{j=1}^{J+1} \prod_{k<j} \pi_{ik} \prod_{k \geq j}(1 - \pi_{ik})}$$

for $j = 1, \ldots, J+1$, where the denominator is a sum over all $J+1$ valid sequences and for notational convenience, $D_0 \equiv 0$ and $D_{J+1} \equiv \infty$. This framework allows us to predict that the MIC is greater than any of the observed dilutions, which is important in detecting emerging resistance.

For a particular model $\mathcal{M}_m$, these probabilities are estimated by plugging in the $\hat{\pi}_{ij}$. To take into account model uncertainty, one can average the above expression over several models using the posterior probabilities of models as weights. While such a procedure may give better estimates, we have not pursued this direction because of the computational demands associated with on-line implementations, and instead use the highest probability model.

The probabilities $\rho_{ij}$ can be used in several ways to determine an estimate of the MIC. We experimented extensively with two approaches: the "modal MIC" and a "decision theoretic MIC." The modal MIC consists simply of choosing the dilution $D_j$ for panel $i$ with the largest probability $\rho_{ij}$. This is the optimal estimator under a loss function that is one if the estimate is not the true MIC and zero if the estimate is exactly the MIC. Our decision theoretic MIC takes into account that under-estimation and over-estimation have different costs and that errors of plus or minus one dilution are unimportant.

4.3. *Decision theoretic MIC estimation.* In evaluating performance in practice, a number of error types are evaluated. The essential agreement between an estimated MIC and reference MIC is defined as agreement in MIC to within $\pm$ one two-fold dilution. According to the Food and Drug Administration (FDA) [FDA (2007), page 21], the essential agreement should be greater than or equal to 90%. In general, underestimation of the MIC by more than one dilution is considered to be worse than overestimating the MIC by more than one dilution.

A second approach for evaluating performance utilizes the categorical classification of isolates as susceptible, intermediate or resistant (SIR). The SIR interpretations are compared and categorical agreement is evaluated. Based on FDA guidelines, the overall categorical agreement should be greater than or equal to 90%. Acceptable error rates are based on the clinical significance of the error. A very major error occurs when an isolate is called susceptible when the isolate is in fact resistant. A very major error rate greater than 1.5% of the resistant strains would be unacceptable. The major error threshold is more forgiving, less than or equal to 3%, and occurs when an isolate is called resistant when in actuality the isolate is susceptible. We devise a decision theoretic estimator that captures the ideas behind the categorical errors, where underestimating the MIC leads to more severe consequences than overestimating it. One way to capture this is to attach a weight, or a loss, to each dilution error size. We use the following notation:

| Error | Loss |
|---|---|
| Exact prediction (no error) | 0 |
| Predicted MIC within 1 dilution of reference MIC | $w_3$ |
| Predicted MIC > reference MIC by more than 1 dilution | $w_2$ |
| Predicted MIC < reference MIC by more than 1 dilution | $w_1$ |

The $w$'s need to be chosen to reflect the relative importance of the errors and will typically be such that $w_1 > w_2 > w_3$.

To obtain the MIC prediction, we choose the dilution that minimizes the expected loss. For dilution $D_j$ this is

(4.2) $$\mathcal{L}_j = w_1 \sum_{k>j+1} \rho_{ik} + w_2 \sum_{k<j-1} \rho_{ik} + w_3(\rho_{i(j-1)} + \rho_{i(j+1)}).$$

The decision theoretic MIC prediction is the $D_j$ associated with the $j$ that minimizes $\mathcal{L}_j$. If one chooses $w_1 = w_2 = w_3$, the modal MIC and the decision theoretic MIC coincide. For the analyses reported later we used $w_1 = 5, w_2 = 1$ and $w_3 = 0$, reflecting that underestimating the MIC by more than one dilution is 5 times worse than overestimating it by more than one dilution. There is no penalty for being within one dilution of the reference MIC.

4.4. *Uncertainty and making a call.* The distribution of the MIC may be useful in formulating guidelines for deciding when to make a call on the MIC prediction or when to continue incubating a panel. One could determine the modal MIC or decision theoretic MIC and how much probability they receive, as well as how much probability mass the estimated MIC plus or minus 1 dilution receives. Sampling could continue until these reach a specified level. We expect that, as sampling continues, the estimated $\pi_{ij}$ will become closer to 1 or 0, so that the probability of the model MIC (the dilution with the largest probability) should increase to 1. However, early on the probabilities will not be as concentrated.

4.5. *Illustration.* Figure 2 shows the MIC distribution for the three bacterial strains that were presented in the growth curves in Figure 1, while Table 2 lists the estimates of the MIC and the probability mass that they receive.

For strain 14,482, the distribution of the MIC is uni-modal with a peak at two, which corresponds to the modal MIC, the decision theoretic MIC and the reference MIC. For strain 14,598, we also have a uni-modal distribution, but underestimate the MIC by one dilution using the modal MIC (reference MIC = 2, modal MIC = 1), while the decision theoretic MIC agrees with the reference MIC. Because the estimate is within one dilution of the reference MIC, this error is not important for essential agreement. Finally, for strain 14,617, the modal MIC is 128, while the reference MIC is 4. While overestimation of the MIC is better than underestimation, this results in an unacceptably large error. Verification of the reference data indicated that the response variable is in fact not in error, but that the problem is likely with the particular panel and/or isolate. The reference MIC and dilutions within one of it receive virtually no support under the estimated distribution of the MIC for strain 14,617, however, for strains 14,482 and 14,598, the probability that the MIC is within one dilution of the reference MIC is 0.96 and 0.97 respectively. Also of interest to note is that for strains 14,482 and 14,598, although the predictions are in agreement with the reference MIC, the probability of the growth sequence being valid is low compared to strain 14,617. While the growth patterns may deviate from the valid order as defined by the respective drug dilutions, this probability does not necessarily provide a reliable indication of the accuracy of the estimated MIC.

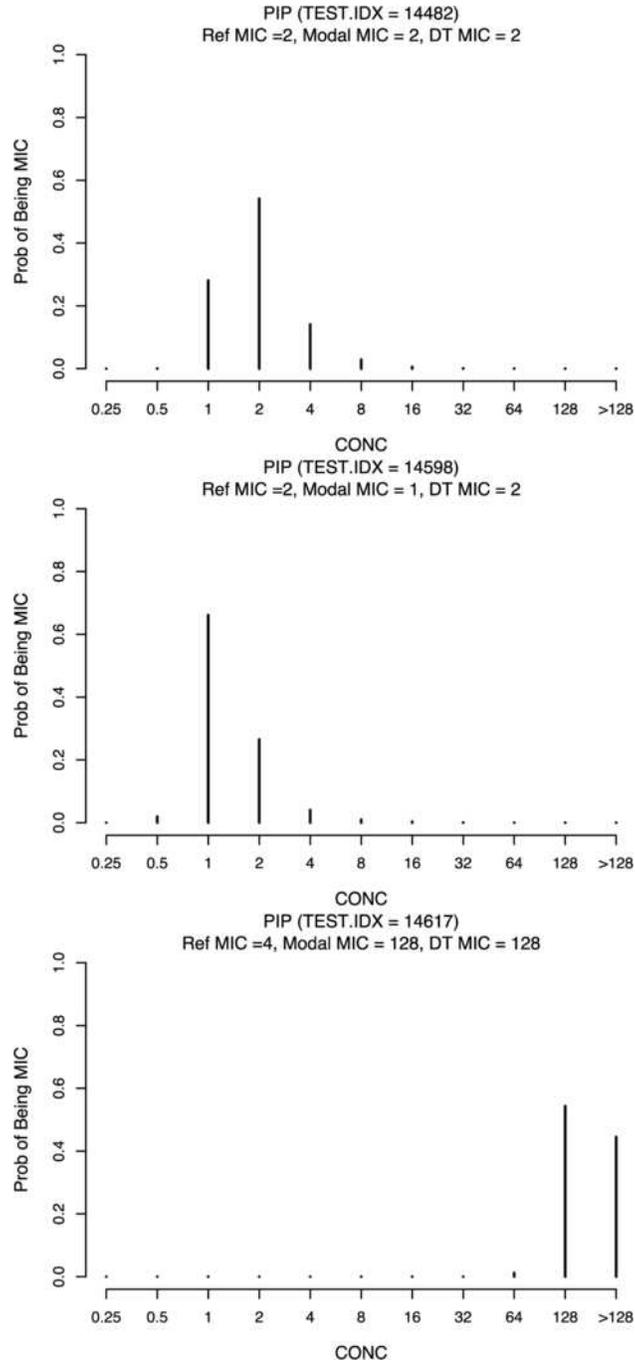

Fig. 2. *Probability distribution of the MIC for the three strains presented in Figure 1.*

TABLE 2
*MIC prediction results for the three strains of interest.* P*(*REF $\pm$ *1) represents the probability of the mass assigned to dilutions within 1 two-fold dilution of the reference MIC.* P(modal MIC) *and* P(DT MIC) *are the probabilities that the modal and the decision theoretic MICs estimates are the MIC, respectively.* P(valid sequence) *is the sum of probabilities of all possible valid sequences*

| Test ID | REF MIC<br>P(REF $\pm$ 1) | Modal MIC<br>P(modal MIC) | Decision theoretic MIC<br>P(DT MIC) | P(valid sequence) |
|---|---|---|---|---|
| 14,482 | 2 | 2 | 2 | 0.33 |
|  | (0.96) | (0.54) | (0.54) |  |
| 14,598 | 2 | 1 | 2 | 0.27 |
|  | (0.97) | (0.66) | (0.26) |  |
| 14,617 | 4 | 128 | 128 | 0.85 |
|  | (0.00) | (0.54) | (0.54) |  |

**5. Validation.** To carry out model validation for each drug, we divide the data into a training and validation group. We used a random sample of 65% of the panel data for training and the remaining 35% for validation. In the training stage we identify the best model using BIC as described previously. This model is then used to predict the MICs for the validation group.

The MIC prediction for the validation group proceeds as follows:

1. The features from each dilution $j$ of each panel $i$ in the validation group are used along with the estimated coefficients from the training data to estimate the probability of growth, $\pi_{ij}$, for dilution $j$, $\hat{\pi}_{ij}$.
2. The probability that a dilution is the MIC for panel $i$, $\hat{\rho}_{ij}$ is calculated [see equation (4.1)]. These probabilities are estimated by plugging in the estimates of $\pi_{ij}$ obtained in Step 1 for each panel in the validation set.

3. The estimated MIC probabilities ($\hat{\rho}_{ij}$, $j = 1, \ldots, J+1$) are then used to estimate the modal or decision theoretic MIC for panel $i$.

For each panel in the validation set we thus obtain a modal and decision theoretic MIC estimator.

5.1. *Essential agreements.* The essential agreements for the FOX nd PIP data are summarized graphically in Figures 3 and 4 and in Tables 3 and 4. In Figures 3 and 4, the choice of MIC estimator affects the error rates. For the modal MIC in FOX, the error rate associated with underestimation by more than one two-fold dilution was 3.82%, whereas for the decision theoretic MIC, the underestimation error rate was 1.56%. Essential agreements for the two MIC estimates were similar, 93.23% for the modal MIC and 93.40% for the decision theoretic MIC. The error rates for PIP are noticeably higher, as anticipated, based on initial perceptions of PIP being a more challenging system. There is a greater tendency for the modal MIC estimate to underestimate the MIC with PIP, than what was observed with FOX. For PIP, the decision theoretic MIC has a much higher essential agreement than the modal MIC, with a reduction by half in the underestimation error rate (12.14% versus 23.57%).

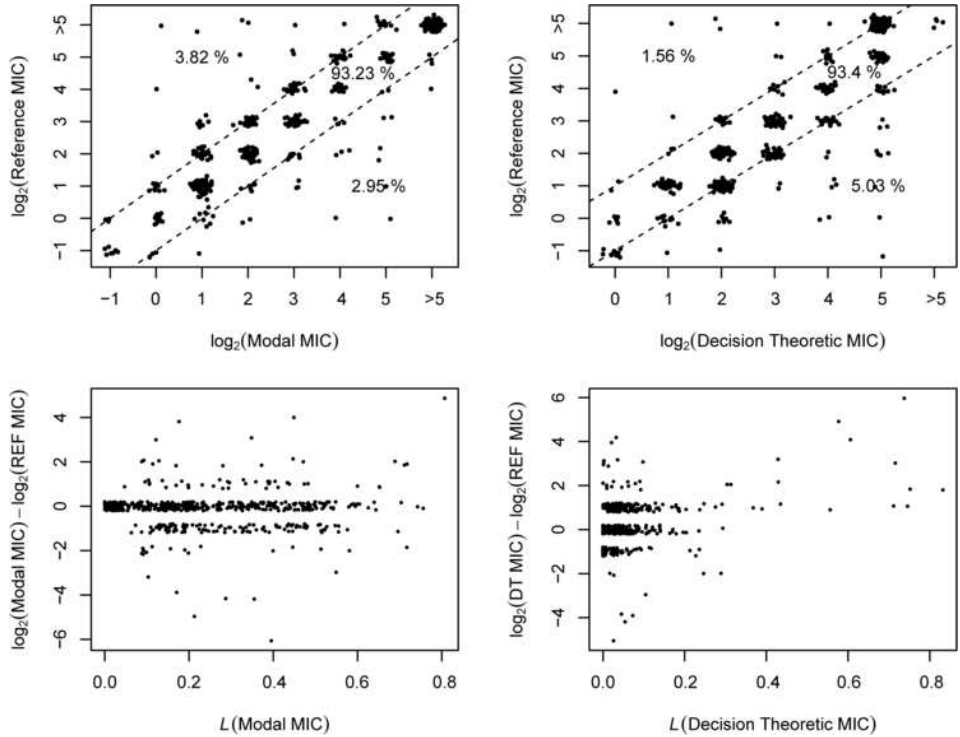

FIG. 3. *Essential agreement (top) and $\log_2$ residual plots (bottom) for the FOX validation set using the modal MIC (left) and the decision theoretic MIC (right).*

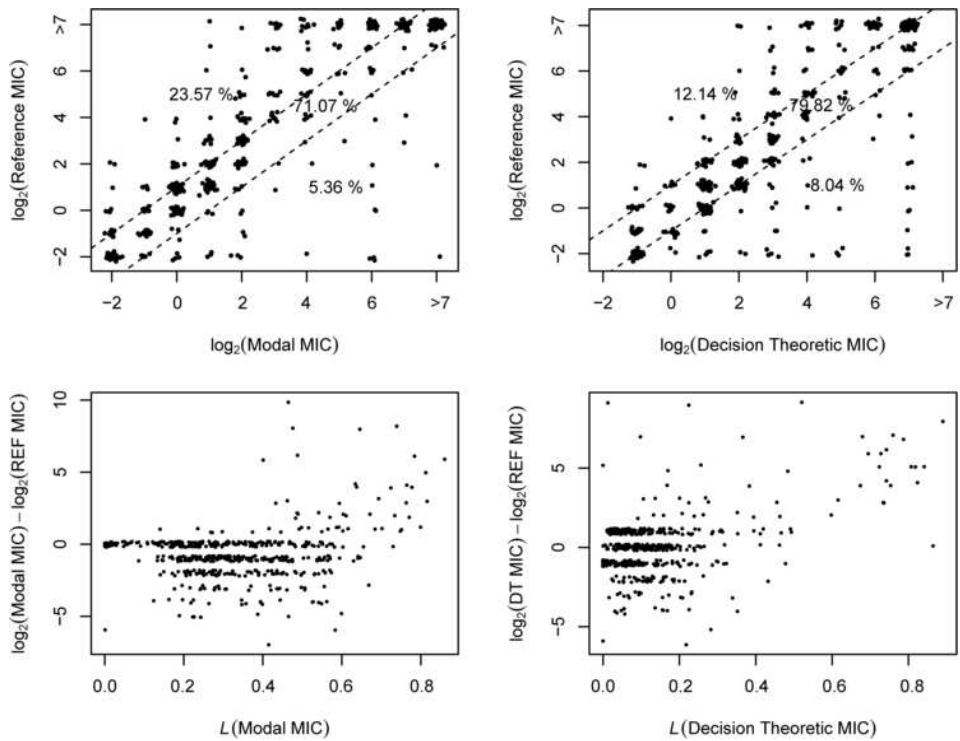

FIG. 4. *Essential agreement (top) and* $\log_2$ *residual plots (bottom) for the PIP validation set using the modal MIC (left) and the decision theoretic MIC (right).*

To further explore the errors, plots of $\log_2$(estimated MIC) $-\log_2$(reference MIC) versus the expected loss evaluated at the estimated MIC are given in Figures 3 and 4, where the estimated MIC is either the modal MIC or decision theoretic MIC. For the modal estimate the corresponding loss is one minus the probability that the modal MIC is the true MIC, $1 - \text{P}(\text{modal MIC})$, and the loss at the decision theoretic MIC, $\mathcal{L}(\text{DT MIC})$, is given by equation (4.2). These plots suggest better agreement between the estimated and reference MIC when the probability of modal MIC is high or loss at the decision theoretic MIC is low. We therefore examined the essential agreements for strains where the probability of the modal MIC was at least 0.5. This leads to a slight improvement for FOX (essential agreement is 94.1% and 94.9% for the modal and decision theoretic estimators, resp.). The gains are greater for PIP with essential agreements of 74.3% and 85.0%, for the modal and decision theoretic estimators, respectively. This suggests that it is possible that by delaying the call because of substantial uncertainty in the MIC and allowing the samples to incubate longer, that we may reduce some of the large errors. This is particularly important for resistant strains that exhibit delayed growth. By delaying the call, the resistance may become more pronounced, leading to correct classification of the curves.

Table 3
*Essential agreements from the FOX validation data*

| Estimate | $\text{EST} - \text{REF} < -1$ | $-1 \leq \text{EST} - \text{REF} \leq -1$ | $\text{EST} - \text{REF} > 1$ |
|---|---|---|---|
| Modal MIC | 3.82 | 93.23 | 2.95 |
| Decision theoretic MIC | 1.56 | 93.40 | 5.03 |

Table 4
*Essential agreements from the PIP validation data*

| Estimate | $\text{EST} - \text{REF} < -1$ | $-1 \leq \text{EST} - \text{REF} \leq -1$ | $\text{EST} - \text{REF} > 1$ |
|---|---|---|---|
| Modal MIC | 23.57 | 71.07 | 5.36 |
| Decision theoretic MIC | 12.14 | 79.82 | 8.04 |

**6. Discussion.** In this paper we have illustrated a probabilistic approach for estimating MICs based on panels of microbial growth curves. Given the necessity to fit a large number of models to hundreds of antibiotics for the implementation in PHOENIX, we had to make several compromises in our modeling approach at the time of algorithm development due to the available computational environments and the computing/storage constraints of the PHOENIX device. Given advances in computing environments today, more flexible models could be obtained by replacing the cubic polynomials with piecewise cubic splines as in generalized additive models. While a fully Bayesian analysis that accounts for errors in the reference MIC might be preferable, the use of logistic regression with approximate model probabilities using BIC and plug-in estimates provided a reasonable solution that could be implemented in real-time in a device with limited computing and storage capacity.

The experimental data presented in this paper to illustrate MIC determination were from the early stages of development of the PHOENIX system. Since our initial analyses were conducted, BD refined formulation of panels and the potencies were adjusted to provide optimum concentrations of antimicrobial agents, leading to a reduction in the bias noted here in PIP. Currently, the methodology described in this paper for the modal MIC is used to make an initial estimate of the MIC in PHOENIX. There are two additional steps in PHOENIX that are used before making a final MIC determination that affect very major and major errors. The first is an expert system that determines if the time is long enough for resistance to express (delayed resistance detection). The second step uses an expert system that takes the results and combines them with other data (resistance markers) to provide a final MIC and SIR determination.

To date, more than seventy-five recent studies have shown the PHOENIX AST system provides accurate estimates of MICs and susceptibility interpretation for various micro-organisms and drugs [see, e.g., Fahr et al. (2003), Donay et al. (2004), Horstkotte et al. (2004)]. Currently, 85 drugs are cleared by the FDA in the U.S. having met FDA standards for essential and categorical agreement, with an additional 20 to 25 in clinical trials. In a European collaborative two-center trial [Fahr et al. (2003)], this system was tested for 469 clinically obtained bacterial isolates with 64 antimicrobial drugs. The results were compared to those of frozen standard broth micro-dilution panels according to the guidelines of the Clinical and Laboratory Standards Institute (formerly the National Committee for Clinical Laboratory Standards). The study [Fahr et al. (2003)] found that performance of the PHOENIX AST system was equivalent to that of the standard broth micro-dilution reference method. In addition to accuracy in prediction, the study [Donay et al. (2004)] found that the PHOENIX AST system required significantly less time to obtain results than by the disk diffusion method.

Predicting the emergence of antibiotic resistance is a challenging problem that many automated AST systems fail to adequately address [Tenover et al. (2006)]. In some cases resistance to higher levels of an antibiotic is virtually unknown. Because the models developed in this paper predict inhibition of growth based on features of the curves, and not actual dilution, the PHOENIX AST system may predict high MICs indicative of emerging resistance, even though the models have not necessarily been trained on resistant strains. For example, the PHOENIX system has already proved to be effective in detecting the emergence of *Staphylococcal* resistance to vancomycin [Deal et al. (2002)].

Wheat, P. F. (2001). History and development of antimicrobial susceptibility testing methodology. *Journal of Antimicrobial Chemotherapy* **48** 1–4.

Xi Kathy Zhou
Department of Public Health
Weill Medical College
Cornell University
New York, New York 10021
USA
E-mail: kaz2004@med.cornell.edu

Merlise A. Clyde
Department of Statistical Science
Duke University
Durham, North Carolina 27708-0251
E-mail: clyde@stat.duke.edu

James Garrett
David Turner
Tim Wiles
Becton–Dickinson
    Diagnostic Systems
Sparks, Maryland 21152
USA
E-mail: Jim_Garrett@bd.com
        David_J_Turner@bd.com
        Tim_Wiles@bd.com

Viridiana Lourdes
Morgan Stanley
New York, New York 10023
USA
E-mail: viridiana.lourdes@morganstanley.com

Michael O'Connell
Waratah Corporation
Durham, North Carolina 27713
USA
E-mail: moconnell@insightful.com

Giovanni Parmigiani
School of Medicine
Johns Hopkins University
Baltimore, Maryland 21205
USA
E-mail: gp@jhu.edu